\documentclass[%
 aip,
 amsmath,amssymb,
 reprint,%
]{revtex4-2}

\usepackage{graphicx}
\usepackage{dcolumn}
\usepackage{bm}

\usepackage[utf8]{inputenc}
\usepackage[T1]{fontenc}
\usepackage{mathptmx}
\usepackage{etoolbox}
\usepackage{xcolor}

\makeatletter
\def\@email#1#2{%
 \endgroup
 \patchcmd{\titleblock@produce}
  {\frontmatter@RRAPformat}
  {\frontmatter@RRAPformat{\produce@RRAP{*#1\href{mailto:#2}{#2}}}\frontmatter@RRAPformat}
  {}{}
}%
\makeatother
\begin{document}

\title{Terahertz third-harmonic generation of lightwave driven Weyl fermions far from equilibrium} 



\author{Patrick Pilch}
\email{patrick.pilch@tu-dortmund.de}
\affiliation{Department of Physics, TU Dortmund University, Dortmund 44227, Germany}
\author{Changqing Zhu}
\affiliation{Department of Physics, TU Dortmund University, Dortmund 44227, Germany}
\author{Sergey Kovalev}
\affiliation{Department of Physics, TU Dortmund University, Dortmund 44227, Germany}
\affiliation{Helmholtz-Zentrum Dresden-Rossendorf, Dresden 01328, Germany}

\author{Renato~M.~A.~Dantas}
\affiliation{Department of Physics, University of Basel, Basel 4056 , Switzerland}
\affiliation{Center of Physics, University of Minho, Braga 4704-553, Portugal}

\author{Amilcar Bedoya-Pinto}
\affiliation{Max Planck Institute for Microstructure Physics, Halle (Saale) 06120, Germany}
\affiliation{Institute of Molecular Science, University of Valencia, Paterna 46980, Spain}
\author{Stuart~S.~P.~Parkin}
\affiliation{Max Planck Institute for Microstructure Physics, Halle (Saale) 06120, Germany}
\author{Zhe~Wang}
\email{zhe.wang@tu-dortmund.de}
\affiliation{Department of Physics, TU Dortmund University, Dortmund 44227, Germany}

\date{\today}

\begin{abstract}
We report on time-resolved ultrafast terahertz third-harmonic generation spectroscopy of nonequilibrium dynamics of Weyl fermions in a nanometer thin film of the Weyl semimetal TaP.  Terahertz third-harmonic generation is observed at room temperature under the drive of a multicycle narrowband terahertz pulse with a peak field strength of down to tens of kV/cm. The observed terahertz third-harmonic generation exhibits a perturbative cubic power-law dependence on the terahertz drive.
By varying the polarization of the drive pulse from linear to elliptical, we realize a sensitive tuning of the third harmonic yield.
By carrying out theoretical analysis based on the Boltzmann transport theory, we can properly describe the experimental results and ascribe the observed THz nonlinearity to field-driven kinetics of the Weyl fermions.
\end{abstract}


\maketitle 

The discovery of Weyl fermions \cite{weyl1929gravitation} in Weyl semimetals is a hallmark breakthrough in condensed matter physics \cite{Jia2016,yan2017topological,Armitage18,Ding21}. 
In this peculiar class of materials, Weyl fermions are realized as the low-energy quasiparticles at the Weyl nodes near the Fermi energy \cite{Jia2016,yan2017topological,Armitage18,Ding21},
which are characterized by linear dispersion relation and nontrivial topological properties.
Deriving from the composite excitation of Dirac fermions, the realization of Weyl fermions in solids requires a reduced symmetry such that two linearly dispersing electron bands corresponding to the Weyl nodes are nondegenerate in the reciprocal space.
The very first significant realizations of the Weyl nodes were made in transition-metal monophosphides, such as TaAs and TaP  \cite{XuHasan15,LvDing15,xu2015experimental,xu2016observation}, where centrosymmetry is indeed absent in contrast to the Dirac semimetals (e.g. Na$_3$Bi and Cd$_3$As$_2$).
As a consequence, finite second-order optical susceptibility is allowed in these noncentrosymmetric compounds, which was found even to be unexpectedly large in the mid- or near-infrared range (see e.g. \cite{Wu2017,Ma2017,sirica2022photocurrent}).

Since Weyl fermions are low-energy excitations in Weyl semimetals, the observed second-order susceptibility in the mid- or near-infrared frequency range (typically $\sim 0.1 - 1$~eV in photon energy) is not necessarily directly related to the low-energy dynamics of Weyl fermions, but rather reflects an average response over the host band structure corresponding to the mid- or near-infrared photon energy.
At the same time, intensive theoretical studies have predicted a variety of fascinating nonlinear features associated to the low-energy dynamics of Weyl fermions (see e.g. \cite{Dantas2021Nonperturbative,BhartiDixit22,
Matus22,Nathan22,Avetissian22,li2022high,
avetissian2022high,Bharti23,Medic24,lim2020efficient}), which largely remain to be investigated experimentally. 
In this work, we employ a low-energy spectroscopic method -- time-resolved terahertz (1~THz $\sim$ 4~meV) high-harmonic generation spectroscopy --  to directly study the nonlinear dynamics of Weyl fermions in the well-established Weyl semimetal TaP.

Efficient THz high-harmonic generation has been found recently in various solid-state quantum systems, such as superconductors \cite{Matsunaga14,Yang2019,Chu2020,Kovalev21,Reinhoffer22}, strongly correlated metals \cite{Reinhoffer24}, semiconductors \cite{Zhu2025}, or topological materials (e.g. graphene \cite{Bowlan14,hafez2018extremely,arshad2023terahertz,maleki2024strategies}, topological insulators \cite{kovalev2021terahertz, tielrooij2022milliwatt,Stensberg23}, and a Dirac semimetal \cite{kovalev2020non,Cheng20,germanskiy2022Ellipticity,ZhuPilch2024}).
The understanding of the THz high-harmonic generation in topological materials has been progressively improved by continuous theoretical investigations (see e.g. \cite{hafez2018extremely,kovalev2020non,Cheng20,
lim2020efficient,Dantas2021Nonperturbative,Mao22,Meng22,
germanskiy2022Ellipticity,ZhuPilch2024}).
Starting from a simplified phenomenological thermodynamic model, one could assume a quasi-instantaneous thermalization of the out-of-equilibrium Dirac fermions in graphene driven by strong THz electric field \cite{hafez2018extremely}.
Under this assumption, the oscillation of the THz field leads to quasi-instantaneous variation of temperature, and consequently a time-dependent optical conductivity. Hence this results in a nonlinear dependence of the current density on the THz field, yielding the THz high-harmonic radiation \cite{hafez2018extremely}.
In contrast, a later more detailed analysis of the nonlinear charge transport figured out explicitly the importance of the THz driven nonequilibrium states in high-harmonic generation for graphene \cite{Mao22}. 
Without assuming a quasi-instantaneous thermalization of the Dirac fermions, one can adopt a more generalized Boltzmann transport theory to describe the THz field driven kinetics of the highly nonequilibrium states \cite{lim2020efficient,Dantas2021Nonperturbative,Meng22}, which has successfully interpreted the experimentally observed THz high-harmonic generation in a three-dimensional Dirac semimetal Cd$_3$As$_2$ \cite{kovalev2020non,Cheng20,
germanskiy2022Ellipticity}.

\begin{figure*}[htb!]
\includegraphics[width=0.9\linewidth]{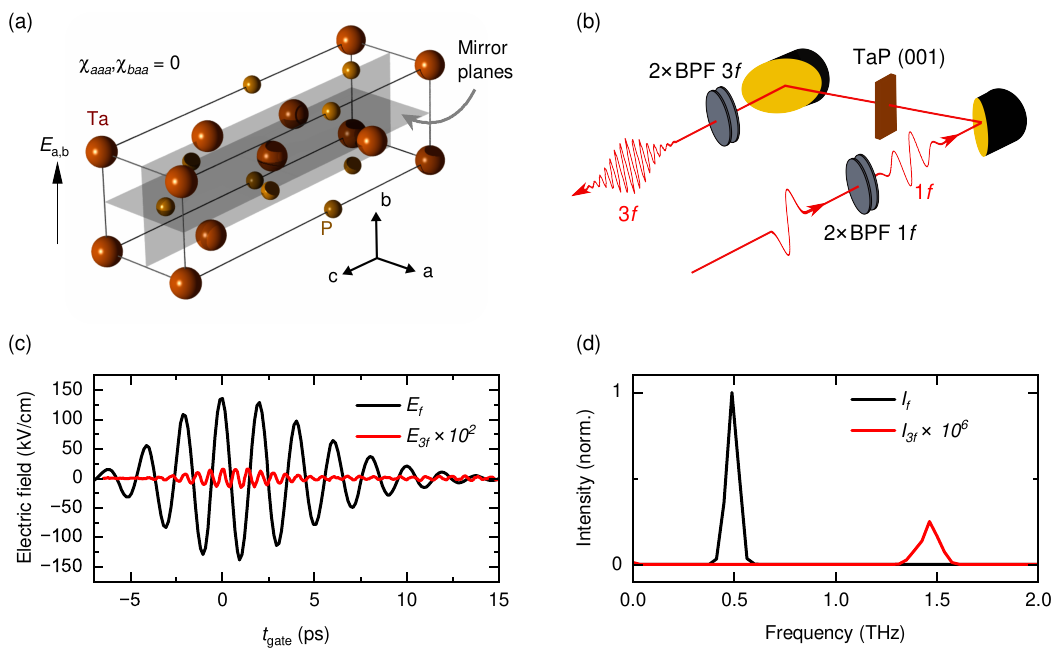}
\caption{
(a) Crystal structure of TaP, with two mirror planes indicated by the grey planes. The THz electric field is parallel to the surface of a (001)-oriented thin film sample plane.
(b) Sketch of the time-resolved THz third-harmonic generation spectroscopy. Two band pass filters (BPF) with central frequency of $1f=0.5$~THz and two with $3f=1.5$~THz are used for preparing a narrowband driving pulse and detection of THG, respectively. 
(c) Time-domain traces of the driving pulse and the third-harmonic emission from the sample, with the frequency-domain Fourier spectra of the time-domain signals in (d).}
\label{Figure1}
\end{figure*}

In a simplified single-particle picture, under the periodic drive of a strong THz electric field, a massless quasiparticle characterized by the Fermi velocity $+ v_F$ or $-v_F$ will periodically pass the Dirac/Weyl node and reverse its velocity, which corresponds to an abrupt change of the electric-current direction in the time domain.
This leads to a very efficient emission of high-order harmonic radiation in the frequency domain.
In a real Dirac/Weyl material, the transport of massless fermions experiences a  variety of scattering processes and is essentially governed by many-body interactions. 
Moreover, the presence of topologically trivial parabolic bands at the Fermi energy can further reduce the nonlinear responses, because the associated charges with a parabolic dispersion relation $\varepsilon \propto k^2$ will behave like free particles that do not contribute to the nonlinear current but mainly absorb the THz field.
Hence, an experimental observation of the terahertz high-harmonic radiation is very challenging. So far, to the best of our knowledge, THz high-harmonic generation still has not been reported in a Weyl semimetal. An experimental study of the THz nonlinear responses
of a Weyl semimetal will not only offer insight into the underlying nonequilibrium many-body physics, but can also provide candidate materials for next generation communication
technologies.

We choose to investigate THz third-harmonic generation in the prototypical Weyl semimetal TaP \cite{xu2015experimental, lee2015fermi, xu2016observation, liu2016evolution}. 
For this study, high quality single-crystalline TaP (001) thin films with a thickness of 22~nm were epitaxially grown on MgO (100) substrates by molecular-beam epitaxy.
A detailed presentation on the sample characterization and the electronic structure fingeprints of Weyl semimetals can be found in Ref.~\cite{bedoya2020realization}.
Although the crystal structure of TaP with a space group $I4_1md$ is noncentrosymmetric, for our (001) oriented film
the elements in the second-order optical susceptibility tensor involving only the $a$ and $b$ axes vanish, due to two mirror planes perpendicular to the crystallographic $a$ and $b$ axes (e.g. $\chi_{aaa}, \chi_{baa} = 0$), see Fig.~\ref{Figure1}(a) for illustration.

Therefore, for our THz electric field parallel to the sample surface, the nonzero second-order nonlinear responses are not involved in our measurements, hence we focus here on the THz third-order harmonic generation.
An intense THz pulse is prepared based on an 800~nm femtosecond laser system by employing LiNbO$_3$ in a tilted pulse-front scheme.
Bandpass filters (BPF) with central frequencies of $f$ or $3f$ and a 20\% bandwidth (FWHM) are utilized to obtain a narrow band THz drive pulse or to filter out the third-harmonic generation (THG) signal, respectively. 
The experimental setup of the time-resolved THz third harmonic generation spectroscopy is sketched in Fig.~\ref{Figure1}(b). 
The THz electric field is gate-detected with femtosecond resolution by electro-optic sampling in a ZnTe crystal \cite{Planken01,Zhang95}.
 
Under the drive of an $f=0.5$~THz pulse, the emission of the nanometer TaP thin film is directly recorded in the time domain after a $3f$-bandpass filter.
The time traces of the driving pulse and the emitted THG electric field are presented in Fig.~\ref{Figure1}(c).
The corresponding spectra in the frequency domain are obtained by Fourier transformation and shown in Fig.~\ref{Figure1}(d), where a THG peak is clearly resolved at $3f=1.5$~THz.
For a maximum peak field value of $E_0=~135$~kV/cm,
we can estimate a field conversion efficiency of $E^\text{peak}_{3f}/E^\text{peak}_f= 0.12\%$, which is smaller in comparison with a reference sample of the topological insulator Bi$_2$Se$_3$ ($\sim 2.9\%$, see Ref.~\cite{tielrooij2022milliwatt}).
Although the efficiency of the THz THG in TaP is not as extremely strong as in Dirac semimetals, the THz THG is unambiguously resolved in our experiment and our result provides a first demonstration of THz THG in a Weyl semimetal.
Various approaches can be utilized to enhance the THz third-harmonic yield, such as tuning the material parameters following our theoretical analysis below [see Eq.~(\ref{THG_Intensity})] or by fabricating THz metamaterials (see e.g. \cite{duan2025arXiv}). 

\begin{figure}[t!]
\includegraphics[width=1\linewidth]{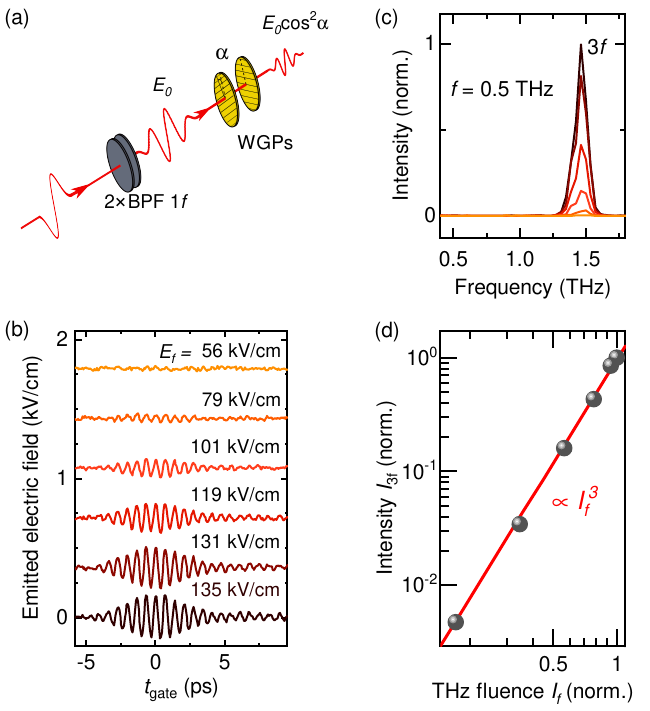}
\caption{(a) Sketch of tuning THz drive field strength by using two wire-grid polarizers (WGPs). The field strength varies with $E_0\cos^2\alpha$, where $\alpha$ represents the polarizer angle, without changing the field polarization. 
(b) Emitted THz field is recorded through a $3f$ bandpass filter in the time domain, for various peak field strengths of an $f=0.5$~THz drive.
(c) Frequency-domain spectra derived through Fourier transformation, exhibiting an evident driving field dependence of the THG intensity.
(d) Fluence dependece of the integrated THG intensity follows a cubic power-law dependence i.e. $I_{3f} \propto I_f^3$ (solid line).}
\label{Figure2}
\end{figure}

In the following, we explore how the nonlinear THz response varies with the driving pulse intensity and ellipticity. 
This will not only demonstrate the possibility to tune the third-harmonic generation of Weyl fermions, but also provide important information for our understanding of the nonlinear nonequilibrium dynamical processes.
In order to study the fluence dependence of the THz THG, we tune the strength of the driving THz field without changing its polarization, by using two THz wire-grid polarizers [see Fig.~\ref{Figure2}(a)].
As shown in Fig.~\ref{Figure2}(b), the emitted THz electric field decreases very rapidly with the decrease of the driving field strength.
Below a peak driving field of $\sim$~70~kV/cm the THG signal can hardly be resolved within the experimental uncertainties. 
These behaviors can also be seen from the corresponding frequency-domain spectra presented in Fig.~\ref{Figure2}(c).
The intensity of the third-harmonic radiation $I_{3f}$ is plotted versus the driving pulse intensity $I_f$ in Fig.~\ref{Figure2}(d).
As shown by the solid line in the double-logarithmic scale, the THG clearly exhibits a power-law dependence, i.e. $I_{3f} \propto I^3_f$.
This follows the expectation for a perturbative response, where an integer $n^\text{th}$-order nonlinear perturbative response $I_{nf}$ is proportional to $I^n_{f}$. 
In contrast, in the Dirac semimetal Cd$_3$As$_2$, non-perturbative THz high-harmonic generation was observed, which does not follow the cubic power-law dependence \cite{kovalev2020non}.
It can be theoretically shown \cite{Dantas2021Nonperturbative,lim2020efficient} that for Dirac and Weyl semimetals, the response remains perturbative as long as the momentum acquired by the quasiparticles due to the driving electric field, $e\,\Delta(t,0)$, does not exceed the Fermi momentum $p_F=\mu/v_F$, i.e., $e\,\Delta(t,0)\le \mu/v_F$. When $e\,\Delta(t,0)>\mu/v_F$, the system crosses into the non-perturbative regime. Here $e\,\bm{\Delta}(t,0)=-e\!\int_{0}^{t}\bm{E}(s)\,ds$, where $e$ is the elementary charge, $v_F$ is the Fermi velocity, $\mu$ is the chemical potential, and $\bm{E}(t)$ is the driving electric field.
This is consistent with the intuition that only when the THz field is sufficiently large, the Weyl system will be driven into the non-perturbative response regime. 
As also will be shown below [see Eq.~(\ref{THG_Intensity})], the efficiency of the harmonic generation is determined by several parameters, including charge carrier density, Fermi velocity, and chemical potential.

\begin{figure}[tb!]
\includegraphics[width=1\linewidth]{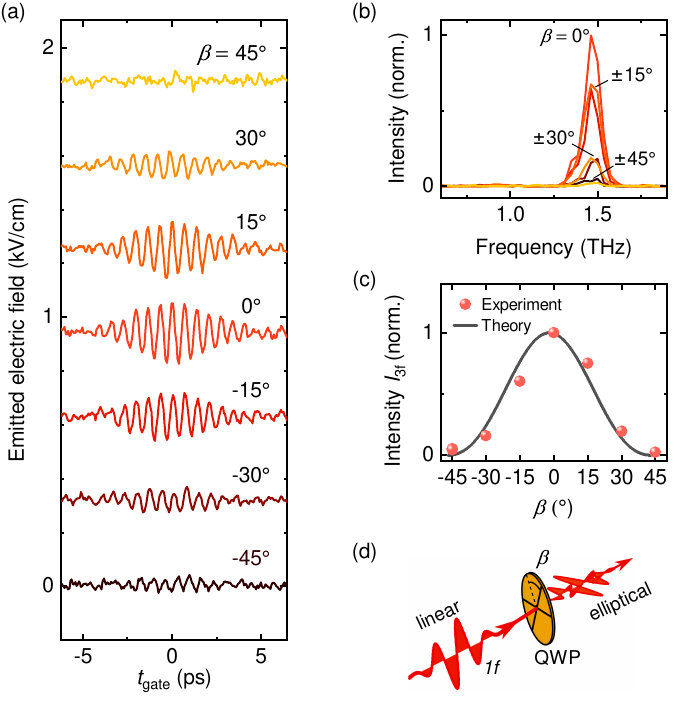}
\caption{
(a) Time-domain signal of the emitted third harmonic field $E_{3f}$ for $f=0.5$~THz at various angles $\beta$ of the THz quarter wave plate (QWP). 
$\beta=0^\circ$ corresponds to linear polarization. For elliptical polarization the dominant parallel component $E_\parallel$ is recorded.
(b) The corresponding frequency-domain spectra derived through Fourier transformation.
(c) Integrated THG intensity $I_{3f,\parallel}$ versus $\beta$. 
(d) Schematic illustration of the driving pulse ellipticity tunable from an incoming linearly polarized THz wave by a QWP.}
\label{Figure3}
\end{figure}

The dependence of high-harmonic generation on driving pulse ellipticity has been extensively investigated theoretically for various theoretical scenarios \cite{TancogneDejean2017,Yoshikawa17,Taucer17,Luo19,Rubio2021,Dantas2021Nonperturbative} and also for high-energy optical excitation in different solid-state materials (see e.g. \cite{Liu2017,You2017,Yoshikawa17}).
To investigate the ellipticity dependence of the THz THG, we employ a commercial quartz quarter-wave plate (QWP) optimized for 0.5~THz to change the ellipticity of the narrowband THz drive pulse.
As illustrated in Fig.~\ref{Figure3}(d), by tuning the angle $\beta$ between the linear polarization direction of the THz drive pulse and the slow/fast axes of the QWP, the obtained THz polarization varies from linear ($\beta=0$) to elliptical ($\beta\neq0$) and almost circular ($\beta=\pm45^\circ$).
It is noteworthy that at $\beta=\pm45^\circ$ the THz polarization is not exactly circular, because the THz drive pulse has a finite bandwidth which experiences dispersion through the QWP. 
Moreover, this dispersion can also result in additional phase shift of the drive pulses, when the polarization is altered from linear to elliptical, which brings in further phase uncertainties.

Figure~\ref{Figure3}(a) displays the experimentally detected electric field of the third-harmonic emission for various polarizations as labeled by the corresponding $\beta$ values, where the data were recorded after a $3f=1.5$~THz bandpass filter. The corresponding frequency-domain spectra obtained through Fourier transformation are presented in Fig.~\ref{Figure3}(b).
For the linearly polarized THz pulse (i.e. $\beta=0^\circ$), a maximum emission of third-harmonic electric field is observed.
With increasing ellipticity the THG efficiency reduces monotonically for either left- or right-handed polarizations,  and reaches a minimum at $\beta=\pm 45^\circ$.
By performing Fourier transformation on the time-domain data, we obtain the THG spectra in the frequency domain, and derive the corresponding THG intensity $I_{3f}$ as integrated over the peak area at $3f$ for different $\beta$ angles.
The ellipticity dependence results are summarized in Fig.~\ref{Figure3}(c). It shows that the THz third-harmonic yield can be efficiently tuned by varying the ellipticity of the THz drive pulse.
A subtle difference between the left- and right-elliptically polarized driving pulses is observed, but comparable to the experimental uncertainties.
We should note that for elliptically or circularly polarized third-harmonic emission, the field component parallel to the initial linear polarization is dominant \cite{germanskiy2022Ellipticity}, which is sensitively detected by our electro-optic sampling scheme \cite{Planken01}.

We can understand the observed ellipticity dependence by the THz field-driven intraband kinetics of Weyl fermions based on the Boltzmann transport theory \cite{Dantas2021Nonperturbative,germanskiy2022Ellipticity}.
Within the relaxation-time approximation the Boltzmann equation reads
\begin{equation}
\partial_t f(t,\bm{p}) - e \bm{E}(t) \cdot \nabla_{\bm{p}} f(t, \bm{p}) = \frac{f_0 (\bm{p}) - f(t,\bm{p})}{\tau}, 
\end{equation}
where $f_0(\bm{p})$ and $f(t,\bm{p})$ are the equilibrium (Fermi-Dirac) and nonequilibrium distribution functions, respectively, and $\tau$ is the relaxation time, treated here as a phenomenological parameter. 
The Boltzmann equation can be solved analytically \cite{Dantas2021Nonperturbative,lim2020efficient,germanskiy2022Ellipticity} to obtain the response current under the THz drive.
For a short relaxation time or a long drive satisfying $\omega \tau \ll 1$, the obtained response current can be written as 
\begin{align}
\bm{j} (t) \approx  \frac{-e \, v_F n }{1-e^{-T/\tau}} \int_{-T}^{0} \frac{du}{\tau}  \,e^{u/\tau} \mathcal{F} \left(\frac{v_F e \bm{\Delta}(t,t+u)}{\mu} \right),
\end{align} 
where $n= \tfrac{\mu^3}{6 \pi^2 \hbar^3 v_F^3}$ is the carrier density and
\[
\mathcal{F} \left(\bm{x}\right) =
\begin{cases}
(1 -\tfrac{1}{5} x^2) \, \bm{x}, &  x \le 1, \\[4pt]
(\tfrac{1}{x} - \tfrac{1}{5} \tfrac{1}{x^3}) \, \bm{x}, & x > 1.
\end{cases}
\] 
The experimentally observed perturbative response of TaP [see Fig.~\ref{Figure2}(d)] corresponds to $ e \Delta(t,t+u)\le \tfrac{\mu}{v_F}$ \cite{Dantas2021Nonperturbative}.
In this regime, for an ideal elliptically polarized pulse $\bm{E} (t)= \mathcal{E} [ \varepsilon_f \cos{(\omega t)} \, \hat{\bm{x}} + \varepsilon_s \sin{(\omega t)} \,\hat{\bm{y}}]$
we can obtain the emitted third-harmonic intensity $\mathcal{I}^{(3 \omega)}_x$ and $\mathcal{I}^{(3 \omega)}_y$ of the $x$ and $y$ components \cite{germanskiy2022Ellipticity}
\begin{align}\label{THG_Intensity}
\left\{ \mathcal{I}^{(3 \omega)}_x, \mathcal{I}^{(3 \omega)}_y \right\} \propto \frac{81 n^2 v_F^8 e^8 \mathcal{E}^6 \omega^2 \tau^6 (\varepsilon^2_f - \varepsilon^2_s)^2 \left\{ \varepsilon^2_f, \varepsilon^2_s\right\}}{100 \, \mu^6 (1+14 \, \omega^2 \tau^2 +49 \, \omega^4 \tau^4 + 36 \, \omega^6 \tau^6)} .
\end{align}
This result means that a maximum third-harmonic emission is achieved for a linearly polarized light (i.e. $\varepsilon_f = 1$ and  $\varepsilon_s = 0$), whereas for circularly polarized light (i.e. $\varepsilon_f = \varepsilon_s$) no third-harmonic generation occurs.
As shown in Fig.~\ref{Figure3}(c), we use Eq.~(\ref{THG_Intensity}) to describe our experimental data very well, confirming the nonlinear dynamical responses of the terahertz field-driven Weyl fermions.
The corresponding mechanism can be delineated as follows. 
Under the drive of a linearly polarized THz electric field, the distribution function of the Weyl fermions is strongly stretched back and forth in the Weyl cones, which deviates far from a Fermi-Dirac distribution for thermal equilibrium.
Due to the linear dispersion relation, this leads to very efficient high-harmonic generation.
For the other limit of a circularly polarized pulse, the shape of the distribution function of the Weyl fermions remains essentially unchanged, hence no harmonic radiation will be generated.
For a finite ellipticity the efficiency of harmonic generation changes monotonically as quantitatively described by Eq.~(\ref{THG_Intensity}) \cite{germanskiy2022Ellipticity}.

In conclusion, by performing time-resolved terahertz third-harmonic generation spectroscopy, we have investigated strong nonlinear responses of Weyl fermions driven by lightwaves far from thermal equilibrium in the Weyl semimetal TaP. 
In particular, we have observed efficient terahertz third-harmonic generation due to the nonlinear dynamics of field-driven Weyl fermions.
The nonlinear dynamics is found to be sensitive to the ellipticity of the terahertz drive pulse. 
Our study demonstrates that in addition to Dirac semimetals and topological insulators, very efficient and tunable terahertz nonlinear responses can also be achieved in Weyl semimetals at room temperature, which implies rich possibilities to explore and utilize terahertz optoelectronic functionalities.

\begin{acknowledgements}

We thank Ahmed Ghalgaoui, Thales V. A. G. de Oliveira, and Yang Zhang for useful discussions.
We acknowledge support by the European Research Council (ERC) under the Horizon 2020 research and innovation programme, grant agreement No. 950560 (DynaQuanta).
A.B.-P. acknowledges support by the Generalitat Valenciana (CIDEGENT/2021/005).\\

P.P. and C.Z. contributed equally to this work.
\end{acknowledgements}


\bibliography{TaP}

\end{document}